\documentclass[12pt,epsfig,color]{article}
\usepackage{amsmath}
\usepackage{epsfig}
\usepackage{amssymb}
\input{epsf}
\newcommand\as{\alpha_{\mathrm{S}}}
\newcommand\f[2]{\frac{#1}{#2}}
\def\la{\lambda}

\def\beq{\begin{equation}}
\def\eeq{\end{equation}}
\def\beeq{\begin{eqnarray}}
\def\eeeq{\end{eqnarray}}
\def\to{\rightarrow}
\def\nn{\nonumber}

\def\b0{b_0}
\def\bone{b_1}

\def\GE{\gamma_E}

\begin{document}

\begin{titlepage}
\renewcommand{\thefootnote}{\fnsymbol{footnote}}
\begin{flushright}
BNL-NT-04/41 \\
RBRC-482 \\
hep-ph/0501258
     \end{flushright}
\par \vspace{10mm}
\begin{center}
{\Large \bf
Threshold Resummation for the Inclusive-Hadron \\[5mm]
Cross-Section in pp Collisions}

\end{center}
\par \vspace{2mm}
\begin{center}
{\bf Daniel de Florian${}^{\,a}$}
\hskip .2cm
and
\hskip .2cm
{\bf Werner Vogelsang$
{}^{\,b}$}\\

\vspace{5mm}
${}^{a}\,$Departamento de F\'\i sica, FCEYN, Universidad de Buenos Aires,\\
(1428) Pabell\'on 1 Ciudad Universitaria, Capital Federal, Argentina

${}^{b}\,$Physics Department and RIKEN-BNL Research Center, \\
Brookhaven National Laboratory, Upton, NY 11973, U.S.A.\\

\end{center}

\par \vspace{9mm}
\begin{center} {\large \bf Abstract} \end{center}
\begin{quote}
\pretolerance 10000
We study the resummation of large logarithmic perturbative corrections
to the partonic cross sections relevant for the process
$pp\to h X$ at high transverse momentum of the hadron $h$.
These corrections arise near the threshold for the partonic
reaction and are associated with soft-gluon emission. We perform
the resummation to next-to-leading logarithmic accuracy.
We present numerical results for the fixed-target regime
and find enhancements over the next-to-leading order
cross section, which significantly improve the
agreement between theoretical predictions and data.
We also apply the resummation for RHIC kinematics and find
that subleading terms appear to play a rather important
role here.

\end{quote}


%
\end{titlepage}

\setcounter{footnote}{1}
\renewcommand{\thefootnote}{\fnsymbol{footnote}}

\section{Introduction}
Cross sections for single-inclusive hadron production
in hadronic collisions, $H_1 H_2\to h X$, play an important role in QCD.
At sufficiently large hadron transverse momentum, $p_T$, one
expects that QCD perturbation theory can be used to derive
predictions for the reaction.
Since high $p_T$ implies
large momentum-transfer, the cross section may be factorized
at leading power in $p_T$ into convolutions of long-distance
pieces representing the structure of the initial hadrons and the
fragmentation of a final-state quark or gluon into the observed
hadron, and parts that are short-distance and describe the hard
interactions of the partons. The long-distance contributions
are universal, i.e., they are the same in any inelastic reaction,
whereas the short-distance pieces depend only on the large scales
related to the large momentum transfer in the overall reaction
and, therefore, can be evaluated using QCD perturbation theory.
Because of this, and because of the fact that single-inclusive
hadrons (e.g., pions) are rather straightforward observables
in experiment, cross sections for $H_1 H_2\to h X$ offer
a variety of important insights into strong interaction dynamics.

If the long-distance pieces, parton distribution functions and
fragmentation functions, are known from other processes, especially
deeply-inelastic scattering and hadron production in
$e^+ e^-$ annihilation, one may test the perturbative
framework outlined above. In particular, one may examine the
relevance of higher orders in the perturbative expansion.
Any discrepancies between the predictions and experimental data
may also provide information about power-suppressed contributions
to the cross section.

Alternatively, one may also gain information
about fragmentation functions. For example, $e^+ e^-$ annihilation
is mostly sensitive to quark-to-hadron fragmentation functions,
whereas data from hadronic collisions may also provide information
on gluon fragmentation. In addition, the reaction $H_1 H_2\to h X$
may be used to probe the structure of the initial hadrons.
Of particular relevance here are spin effects, associated
with polarized initial protons. At the BNL Relativistic Heavy-Ion
Collider (RHIC), one measures spin asymmetries in polarized $pp
\to h X$ scattering, in order to investigate the spin structure
of the nucleon. Finally, high-$p_T$ hadrons are also important
probes of strongly interacting matter in a high-energy nuclear
environment, as generated by heavy-ion collisions. Here, hadron
production in proton-proton collisions provides an important
baseline for the study of nuclear dynamics.

Whatever the uses of processes $H_1 H_2\to h X$, a central piece
is in each case the perturbative partonic hard scattering and our
ability to reliably evaluate it. Lowest-order (LO) calculations
of the partonic short-distance cross sections were performed a
long time ago~\cite{old}, and later improved when the next-to-leading
order (NLO) corrections were computed~\cite{Aversa:1988vb,ddf,Jager:2002xm}.
On the experimental side,
an extensive data set on high-$p_T$ single-inclusive hadron data
has been collected, both from scattering off fixed targets and
from colliders at much higher energies. Detailed comparisons
of NLO calculations with the experimental data have been carried out
recently in~\cite{aur,apan,bs,kkp1}. They show the overall trend
that NLO theory significantly underpredicts the cross section data at
fixed-target energies, but yields a good
description~\cite{kkp1,phenix,star} of the collider data.

In the present paper, we further improve the theoretical
calculations by implementing the all-order resummation of large
logarithmic corrections to the partonic cross sections.
At partonic threshold, when the initial partons have
just enough energy to produce a high-transverse momentum
parton (which subsequently fragments into the observed hadron)
and a massless recoiling jet, the phase space available for gluon
bremsstrahlung vanishes, resulting in large logarithmic corrections to
the partonic cross section. To be more specific, if we consider the
cross section as a function of the hadron transverse momentum $p_T$,
integrated over all hadron rapidity, the partonic threshold
is reached when $\sqrt{\hat{s}}=2 \hat{p}_T$, where $\sqrt{\hat{s}}$
is the partonic center-of-mass (c.m.) energy, and $\hat{p}_T=p_T/z$ is the
transverse momentum of the produced parton fragmenting into
the hadron, the latter taking the fraction $z$ of the
parton momentum. Defining $\hat{x}_T\equiv 2 \hat{p}_T/\sqrt{\hat{s}}$,
the leading large contributions near threshold arise as $\as^k
\ln^{2k}\left(1-\hat{x}_T^2\right)$ at the $k$th order in
perturbation theory, where $\as$ is the strong coupling. Sufficiently
close to threshold, the perturbative series will be only useful if
such terms are taken into account to all orders in $\as$, which is
what is achieved by threshold resummation~\cite{dyresum,KS,BCMN}.
This resummation has been derived for a number of cases of
interest, to next-to-leading logarithmic (NLL) order.
As far as processes with similar kinematics are concerned,
it has been investigated for high-$p_T$ prompt
photon production in hadronic collisions~\cite{LOS,CMN,KO,sv}, and
also for jet production~\cite{KS,KOS,KO1}, which proceeds through
the same partonic channels as inclusive-hadron production. We
will actually make use of the results of~\cite{KOS,KO1} for the resummed
jet cross section in our analysis below.

The larger $\hat{x}_T$, the more dominant the threshold logarithms
will be. Since $\hat{s}=x_1 x_2 S$, where $x_{1,2}$ are the partonic
momentum fractions and $\sqrt{S}$ is the hadronic c.m. energy,
and since the parton distribution functions
fall rapidly with increasing $x_{1,2}$, threshold effects
become more and more relevant as the hadronic scaling variable
$x_T\equiv 2 p_T/\sqrt{S}$ goes to one. This means that
the fixed-target regime with 3~GeV $\lesssim p_T\lesssim$ 10~GeV
and $\sqrt{S}$ of 20$-$30~GeV is the place where threshold
resummations are expected to be particularly relevant and useful.
We will indeed confirm this in our study. Nonetheless, because
of the convoluted form of the partonic cross sections and the
parton distributions and fragmentation functions (see below),
the threshold regime $\hat{x}_T\to 1$ plays an important
role also at much higher (collider) energies. Here one may,
however, also have to pay attention to terms that are subleading
near threshold.

In Sec.~\ref{sec2} we provide the basic formulas for the
inclusive-hadron cross section at fixed order in perturbation
theory, and display the role of the threshold region.
Section~\ref{sec3} presents details of the threshold
resummation for the inclusive-hadron cross section.
In Sec.~\ref{sec4} we give phenomenological
results. We focus primarily on the fixed-target regime, but
also give some exploratory results for collider energies.
We do not present an exhaustive phenomenological analysis
of all hadron production data available, but select some
representative examples. Finally, we summarize our results
in Sec.~\ref{sec5}. The Appendix compiles some useful
formulas for the threshold-resummed cross section.

\section{Perturbative cross section and the threshold region
\label{sec2}}

We consider single-inclusive hadron production in hadronic collisions,
\begin{align}
H_1(P_1) + H_2(P_2) \rightarrow h(P_3) + X \, ,
\end{align}
at large transverse momentum $p_T$ of hadron $h$. We integrate
over all angles (equivalently, pseudorapidities $\eta$) of the produced
hadron. We note from the outset that this does not directly
correspond to the experimental situation where always only a
certain range in pseudorapidity is covered; we will return to
this point later on. The factorized cross section for the process
can then be written in terms of the convolution
\begin{align}
\label{eq:1} \f{p_T^3\, d\sigma(x_T)}{dp_T} = \sum_{a,b,c}\, &
\int_0^1 dx_1 \, f_{a/H_1}\left(x_1,\mu_{FI}^2\right) \, \int_0^1
dx_2 \, f_{b/H_2}\left(x_2,\mu_{FI}^2\right) \, \int_0^1 dz
\,z^2\, D_{h/c}\left(z,\mu_{FF}^2\right) \, \nn \\ &\int_0^1
d\hat{x}_T \, \, \delta\left(\hat{x}_T-\f{x_T}{z\sqrt{x_1
x_2}}\right) \, \int_{\hat{\eta}_{-}}^{\hat{\eta}_{+}} d\hat{\eta}
\, \f{\hat{x}_T^4 \,\hat{s}}{2} \,
 \f{d\hat{\sigma}_{ab\rightarrow cX}(\hat{x}_T^2,\hat{\eta})}{ d\hat{x}_T^2 d\hat{\eta}} \, ,
\end{align}
where $\hat{\eta}$ is the pseudorapidity at parton level, with
$\hat{\eta}_{+}=-\hat{\eta}_{-}=\ln\left[(1+\sqrt{1-\hat{x}_T^2})/
\hat{x}_T\right]$. The sum in Eq.~(\ref{eq:1}) runs over all
partonic subprocesses $ab\to cX$, with partonic cross sections
$d\hat{\sigma}_{ab\rightarrow cX}$, parton distribution functions
$f_{a/H_1}$ and $f_{b/H_2}$, and parton-to-hadron fragmentation
functions $D_{h/c}$. The scales $\mu_{FI}$ and $\mu_{FF}$ denote
the factorization scales for the initial and final states,
respectively. The dependence on them, and on the renormalization
scale $\mu_R$, is implicit in the partonic cross section in
Eq.~(\ref{eq:1}).

The partonic cross sections are computed in QCD perturbation theory.
Their expansions begin at ${\cal O}(\as^2)$ since the LO partonic
processes are the $2\to 2$ reactions $ab\to cd$. Therefore,
\begin{equation}
d\hat{\sigma}_{ab\rightarrow cX}(\hat{x}_T^2,\hat{\eta})=
\as^2(\mu_R)\,\Big[ d\hat{\sigma}_{ab\rightarrow cd}^{(0)}
(\hat{x}_T^2,\hat{\eta}) + \as(\mu_R)\,
d\hat{\sigma}_{ab\rightarrow cX}^{(1)}(\hat{x}_T^2,\hat{\eta}) +
{\cal O}(\as^2) \Big] \, .
\end{equation}
It is customary to express $\hat{x}_T^2$ and $\hat{\eta}$ in
terms of a different set of variables, $v$ and $w$:
\begin{equation}
\label{vwdef}
\hat{x}_T^2=4 v w(1-v) \;\;\;\;\;\;\;\;
{\rm e}^{2\hat{\eta}}=\f{v w}{1-v} \,.
\end{equation}
At LO, one then has
\begin{align}
\label{eq:sigma0} \f{\hat{s}\, d\hat{\sigma}_{ab\rightarrow
cd}^{(0)}(v,w)}{dv\,dw} = \f{\hat{s}\, d\hat{\tilde{\sigma}}_
{ab\rightarrow cd}^{(0)}(v)}{dv}\,\delta(1-w) \; ,
\end{align}
where the $\delta(1-w)$ function simply expresses the fact that
$\hat{x}_T\cosh(\hat{\eta})\equiv 1$ for $2\to 2$ kinematics.
It allows to trivially perform the $\hat{\eta}$ integration
of the partonic cross section. Defining
\begin{equation}
\Sigma_{ab\rightarrow cX}(\hat{x}_T^2)\equiv
\int_{\hat{\eta}_{-}}^{\hat{\eta}_{+}} d\hat{\eta} \,
\f{\hat{x}_T^4 \,\hat{s}}{2} \, \f{d\hat{\sigma}_{ab\rightarrow
cX}(\hat{x}_T^2,\hat{\eta})}{ d\hat{x}_T^2 d\hat{\eta}} \, ,
\end{equation}
the LO cross section for the process $gg\to gg$ becomes, for example,
\begin{equation}
\Sigma_{gg\rightarrow gg}^{(0)}(\hat{x}_T^2)= 18 \as^2
\pi\,\f{\left(1-\frac{\hat{x}_T^2}{4}\right)^3}{
\sqrt{1-\hat{x}_T^2}} \; .
\end{equation}
Analytical expressions for the NLO corrections
$d\hat{\sigma}_{ab\rightarrow cX}^{(1)}(v,w)$ have been obtained
in~\cite{Aversa:1988vb,Jager:2002xm}. Schematically, they read:
\begin{align}
\label{eq:sigma1}
\f{\hat{s}\, d\hat{\sigma}_{ab\rightarrow cX}^{(1)}(v,w)}{dv\,dw}
= A(v)\, \delta(1-w) + B(v)\, \left(\f{\ln(1-w)}{1-w} \right)_+ + C(v)\,
\left(\f{1}{1-w}  \right)_++ F(v,w) \; ,
\end{align}
where the $+$ distributions are defined in the usual way,
\begin{equation}
\int_0^1 f(w)\left[g(w)\right]_+\,dw =\int_0^1 \left[f(w)-f(1)
\right] g(w)\, dw \;\; .
\end{equation}
The function $F(w,v)$ in Eq.~(\ref{eq:sigma1})
represents all remaining terms without distributions in $w$.
The terms with $+$ distributions in Eq.(\ref{eq:sigma1})
generate at NLO level the large logarithmic contributions we
discussed earlier, and which we will resum to all orders in $\as$.
After integration over $\hat{\eta}$, the term $[\ln(1-w)/(1-w)]_+$
yields a contribution $\propto \ln^2 (1-\hat{x}_T^2)$
to $\Sigma_{ab\rightarrow cX}^{(1)}$, plus terms less singular
at $\hat{x}_T=1$. At higher orders, the leading logarithmic
contributions are enhanced by terms proportional to $\as^k \,
[\ln^{2k-1}(1-w)/(1-w)]_+$ in $d\hat{\sigma}_{ab\rightarrow cX}^{(k)}
(v,w)/dv\,dw$, or to $\as^k \, \ln^{2k} (1-\hat{x}_T^2)$
in $\Sigma_{ab\rightarrow cX}^{(k)}$. As we discussed earlier,
these logarithmic terms are due to soft-gluon radiation and,
because there are two additional powers of the logarithm for
each new order in perturbation theory, may spoil the
perturbative expansion unless they are resummed to all orders.

As follows from Eq.~(\ref{eq:1}), since the hadronic
variable $x_T$ is fixed, $\hat{x}_T$ assumes particularly
large values when the partonic momentum fractions approach
the lower ends of their ranges. Since the parton distributions
and fragmentation functions rise steeply towards small argument,
this generally increases the relevance of the threshold
regime and the soft-gluon effects are relevant even for situations
where the the hadronic center-of-mass energy is much larger than the
transverse momentum of the final state hadrons. This effect, valid in
general in hadronic collisions, is even enhanced in single-inclusive
hadron production since only a fraction $z$ of the available energy
is actually used to produce the final-state hadron.

\section{Resummed cross section \label{sec3}}

We will now present the formulas for the threshold-resummed partonic
cross sections. We will do this only for the case of the fully
rapidity-integrated cross section, which turns out to significantly
simplify the analysis. The resummation for the $\eta$ dependence
of the kinematically related prompt-photon cross section was
performed in Ref.~\cite{sv}, and we could in principle follow
the techniques developed there to derive the $\eta$ dependence
of the resummed inclusive-hadron cross section. However, this
process is of much greater complexity than prompt photons, as
will become evident below, and it appears that a successful
resummation at fixed rapidity will require further new techniques.
We hope to address this in a future publication.
As far as phenomenology is concerned, we will later on mimic
the effects of the experimentally covered limited rapidity ranges
by rescaling our resummed prediction by an appropriate
ratio of NLO cross sections. Such an approximation was shown
in~\cite{sv} to work extremely well for the resummed prompt-photon
cross section, where it was found that the shape of the cross
section as a function of rapidity does not change much when
going from NLO to the resummed result. This gives confidence
that it may be applicable also in the case of inclusive-hadron
production we are interested in.

\subsection{Mellin moments and threshold region}

The resummation of the soft-gluon contributions is carried out in
Mellin-$N$ moment space, where the convolutions in Eq.~(\ref{eq:1})
between parton
distributions, fragmentation functions, and subprocess cross sections
factorize into ordinary products. We take Mellin moments in the scaling
variable $x_T^2$ as
\begin{align}
\label{eq:moments}
\sigma(N)\equiv \int_0^1 dx_T^2 \, \left(x_T^2 \right)^{N-1} \;
\f{p_T^3\, d\sigma(x_T)}{dp_T} \, .
\end{align}
In $N$-space Eq.(\ref{eq:1}) becomes
\begin{align}
\sigma(N)=\sum_{a,b,c} \,  f_{a/H_1}(N+1,\mu_{FI}^2) \,
f_{b/H_2}(N+1,\mu_{FI}^2) \,  D_{h/c}(2N+3,\mu_{FF}^2) \,
\hat{\sigma}_{ab\rightarrow cX}(N)\, ,
\end{align}
with the Mellin moments of the parton distribution functions
and fragmentation functions, and where
\begin{align} \label{momdef}
\hat{\sigma}_{ab\rightarrow cX}(N) \equiv \int_0^1 d\hat{x}_T^2 \,
\left(\hat{x}_T^2 \right)^{N-1}\, \Sigma_{ab\rightarrow
cX}(\hat{x}_T^2)= \f{1}{2} \int_0^1 dw  \int_0^1 dv\, \left[4
v(1-v)w\right]^{N+1}\, \f{\hat{s}\, d\hat{\sigma}_{ab\rightarrow
cX}^{(1)}(w,v)}{dw\,dv} \,.
\end{align}
Here, the threshold limit $w\to 1$ (or, for the rapidity-integrated
cross section, $\hat{x}_T^2\to 1$) corresponds to $N\to \infty$,
and the leading soft-gluon corrections arise as terms
$\propto \as^k \ln^{2k}N$.

It is instructive to examine the interplay of rapidity integration
and large-$N$ limit, for example in case of the NLO cross
section in Eq.~(\ref{eq:sigma1}). As we mentioned earlier, the
soft-gluon terms are associated with the $+$ distribution pieces
in~(\ref{eq:sigma1}), which have coefficients that may be
written as functions of $v$ only. One has
\begin{align}
\int_0^1\, dv\,  \left[4 v(1-v)w\right]^{N+1}\, f(v) =
\int_0^1\, dv\,   \left[4 v(1-v)w\right]^{N+1} \left[ f\left(\f{1}{2}\right)
+{\cal O}\left(\f{1}{N}\right) \right]\, ,
\end{align}
which implies that at large $N$ the variable $v$ is ``squeezed'' to
$v=1/2$, and hence, as follows from~(\ref{vwdef}), the partonic
rapidity is forced to $\hat{\eta}=0$. This means, for example,
that near threshold it is justified to take the NLO partonic cross
sections as proportional to the Born cross section,
\begin{align}
\label{eq:sigma1p} \f{\hat{s}\, d\hat{\sigma}_{ab\rightarrow
cX}^{(1)}(v,w)}{dv\,dw} \approx \f{\hat{s}\,
d\hat{\tilde{\sigma}}_ {ab\rightarrow cd}^{(0)}(v)}{dv}\; \left[
A'\, \delta(1-w) + B'\, \left(\f{\ln(1-w)}{1-w} \right)_+ + C'\,
\left(\f{1}{1-w}  \right)_+ \,\right] \; ,
\end{align}
with coefficients $A',B',C'$ evaluated at $v=1/2$. We will
follow this reasoning also for the resummed cross section to which
we turn now.

\subsection{Resummation to NLL}
\label{subsec32}

In Mellin-moment space, threshold resummation results in
exponentiation of the soft-gluon corrections. Foremost,
there are radiative factors for the initial and final
partons, which contain the leading logarithms.
At variance with the color-singlet cases of Drell-Yan
and Higgs production~\cite{dyresum,higgs,higgsnnll}, and with prompt-photon
production~\cite{LOS,CMN} which has only one color structure at Born
level, several color channels contribute to
each of the $2\to 2$ QCD subprocesses relevant for inclusive-hadron
production. As a result, there are color interferences and correlations
in large-angle soft-gluon emission at NLL, and the resummed cross
section for each subprocess becomes a sum of exponentials, rather
than a single one.

In determining the resummed formula, we are in the fortunate
situation that the effects of the color interferences for
soft-gluon emission in the $2\to 2$ processes $ab\to cd$
have been worked out in detail in Refs.~\cite{KOS,KO1} for the
case of jet production in hadronic collisions, which proceeds
through the same Born processes. We take advantage of
the formulas derived there. The results in~\cite{KOS,KO1}
have actually been given for arbitrary rapidity; for the
case of the rapidity-integrated cross section we consider
here it is sufficient to set $\hat{\eta}=0$ in the expressions
of~\cite{KOS,KO1} when diagonalizing (by changing the color basis)
the soft anomalous dimension matrix computed there.
A difference between inclusive
hadrons and jets occurs regarding the treatment of the
final-state parton $c$ producing the jet or the hadron.
In our case, this parton is ``observed'', that is, we
are considering a single-inclusive parton cross section.
Such a cross section has final-state collinear singularities
which are factorized into the fragmentation functions. As
far as resummation is concerned, the final-state observed
parton therefore is similar to the initial-state partons
and receives essentially the same radiative factor as the
latter~\cite{cc}.

Combining these results of~\cite{dyresum,KOS,KO1,cc}, we can cast
the resummed partonic cross section for each subprocess
into the rather simple form
\begin{align}
\label{eq:res}
\hat{\sigma}^{{\rm (res)}}_{ab\to cd} (N)=  C_{ab\to cd}\,
\Delta^a_N\, \Delta^{b}_N\, \Delta^{c}_N\,
J^{d}_N\, \left[ \sum_{I} G^{I}_{ab\to cd}\,
\Delta^{{\rm (int)} ab\rightarrow cd}_{I\, N}\right] \,
\hat{\sigma}^{{\rm (Born)}}_{ab\to cd} (N) \;  ,
\end{align}
where the sum runs over all possible color configurations $I$,
with $G^{I}_{ab\to cd}$ representing a weight for
each color configuration, such that $\sum_I G^{I}_{ab\to cd}=1$.
$\hat{\sigma}^{{\rm (Born)}}_{ab\to cd}(N)$ denotes the $N$-moment
expression for the Born cross section for the process, as defined
in Eq.~(\ref{momdef}). We list the moment space expressions
for all the Born cross sections in the Appendix.
Each of the functions $\Delta^{i}_N$, $J^{d}_N$,
$\Delta^{{\rm (int)} ab\rightarrow cd}_{I\, N}$ in Eq.~(\ref{eq:res})
is an exponential. $\Delta^a_N$ represents the effects of
soft-gluon radiation collinear to initial parton $a$ and is
given, in the $\overline{{\mathrm{MS}}}$ scheme, by
\begin{align}\label{Dfct}
\ln \Delta^a_N&=  \int_0^1 \f{z^{N-1}-1}{1-z}
\int_{\mu_{FI}^2}^{(1-z)^2 Q^2} \f{dq^2}{q^2} A_a(\as(q^2)) \; ,
\end{align}
and similarly for $\Delta^b_N$. Collinear soft-gluon
radiation to parton $c$ yields the same function, but
with the initial-state factorization scale $\mu_{FI}$ replaced
with the final-state one, $\mu_{FF}$. The function $J^{d}_N$
embodies collinear, soft or hard, emission by the non-observed
parton $d$ and reads:
\begin{align} \label{Jfct}
\ln J^d_N&=  \int_0^1 \f{z^{N-1}-1}{1-z} \Big[
\int_{(1-z)^2 Q^2}^{(1-z) Q^2} \f{dq^2}{q^2} A_a(\as(q^2)) +
\f{1}{2} B_a(\as(1-z)Q^2) \Big] \; .
\end{align}
Large-angle soft-gluon emission is accounted for by the factors
$\Delta^{{\rm (int)} ab\rightarrow cd}_{I\, N}$, which depend on
the color configuration $I$ of the participating partons.
Each of the $\Delta^{{\rm (int)} ab\rightarrow cd}_{I\, N}$
is given as
\begin{align}\label{Dintfct}
\ln\Delta^{{\rm (int)} ab\rightarrow cd}_{I\, N} &=
 \int_0^1 \f{z^{N-1}-1}{1-z} D_{I\, ab\to cd}(\as((1-z)^2 Q^2)) \; .
\end{align}
Finally, the coefficient $C_{ab\to cd}$ contains
$N-$independent hard contributions arising from one-loop
virtual corrections.

In the above formulas, Eqs.~(\ref{Dfct})-(\ref{Dintfct}),
we have defined $Q^2=2 p_T^2$. Furthermore, each of the functions
${\cal F}\equiv A_a$, $B_a$, $D_{I\, ab\to cd}$
is a perturbative series in $\as$,
\begin{equation}
{\cal F}(\as)=\frac{\as}{\pi} {\cal F}^{(1)} +
\left( \frac{\as}{\pi}\right)^2 {\cal F}^{(2)} + \ldots \; ,
\end{equation}
with~\cite{KT}:
\begin{equation}
\label{A12coef}
A_a^{(1)}= C_a
\;,\;\;\;\; A_a^{(2)}=\frac{1}{2} \; C_a  \left[
C_A \left( \frac{67}{18} - \frac{\pi^2}{6} \right)
- \frac{5}{9} N_f \right]
\;,\;\;\;\; B_a^{(1)}=\gamma_a \; ,
\end{equation}
where $N_f$ is the number of flavors, and
\begin{eqnarray}
&&C_g=C_A=N_c=3 \;, \;\;\;C_q=C_F=(N_c^2-1)/2N_c=4/3 \nn \\
&&\gamma_q=-3 C_F/2=-2\; , \;\;\; \gamma_g=-2\pi \b0\; , \;\;\;
\b0 = \frac{1}{12 \pi} \left( 11 C_A - 2 N_f \right) \; .
\end{eqnarray}
The expansion of the coefficients $C_{ab\to cd}$ reads:
\begin{eqnarray}
C_{ab\to cd} = 1 + \frac{\as}{\pi} C_{ab\to cd}^{(1)} + {\cal O}(\as^2) \; .
\end{eqnarray}

In the exponents, the large logarithms in $N$
now occur only as single logarithms, of the form
$\as^k \ln^{k+1}(N)$ for the leading terms. Subleading terms
are down by one or more powers of $\ln(N)$. Knowledge of the
coefficients $A_a^{(1,2)}$, $B_a^{(1)}$, $D_{I\, ab\to cd}^{(1)}$
allows to resum the full towers of leading logarithms (LL)
$\as^k \ln^{k+1}(N)$, and NLL $\as^k \ln^k(N)$ in the exponent.
Along with the coefficients $C_{ab\to cd}^{(1)}$ one then gains
control of three towers of logarithms in the cross section,
$\as^k \ln^{2k}(N)$, $\as^k \ln^{2k-1}(N)$, $\as^k \ln^{2k-2}(N)$,
which is likely to lead to a much improved theoretical prediction.
We also note that the factors $\Delta_N^i$ depend on the
initial- or final-state factorization scales in such a way
that they will compensate the scale dependence (evolution)
of the parton distribution and fragmentation functions. One
therefore expects a decrease in scale dependence, which indeed
has been found in previous studies for other threshold-resummed
cross sections.

We finally examine the qualitative impact of the resummation.
To this end, we note that, neglecting the running of the
strong coupling, the LL terms in the exponents in Eqs.~(\ref{Dfct})
and (\ref{Jfct}) become
\begin{eqnarray}\label{DJfct1}
\Delta^a_N&=&  \exp\left[ \frac{\as}{\pi}C_a\ln^2 (N) \right] \; , \nn \\
J^d_N&=& \exp\left[ -\frac{\as}{2\pi}C_d \ln^2 (N) \right]  \; .
\end{eqnarray}
Therefore, for each partonic channel, the leading logarithms
are
\begin{align}
\label{eq:res1}
\hat{\sigma}^{{\rm (res)}}_{ab\to cd} (N) \propto
\exp\left[ \frac{\as}{\pi}\left(C_a+C_b+C_c-\frac{1}{2} C_d\right)
\ln^2 (N) \right] \; .
\end{align}
The fact that this exponent is clearly positive for each of the
partonic channels means that the soft-gluon effects will lead to an
enhancement of the cross section. Particularly strong enhancements
are to be expected for gluonic channels; for example, for
the process $gg\to gg$ one has $C_a+C_b+C_c-C_d/2=15/2$.
The feature that partonic cross sections can give Sudakov
enhancements is
related to the fact that finite partonic cross sections are
obtained after collinear (mass) factorization, so that
soft-gluon effects are partly already contained in the
($\overline{{\mathrm{MS}}}$-defined) parton distribution functions and,
in our case, fragmentation functions.

\subsection{Exponents at NLL}

We now give explicit formulas for the expansions of the resummed
exponents to NLL accuracy. Since the functions $\Delta^{i}_N$ and
$J^{d}_N$ are ``universal'' in the sense that they depend only on
the type of the external parton, but not on the subprocess, their
expansions are known and we recall them for the sake of
completeness: \beeq \label{lndeltams} \!\!\! \!\!\! \!\!\! \!\!\!
\!\!\! \ln \Delta_N^a(\as(\mu_R^2),Q^2/\mu_R^2;Q^2/\mu_F^2)
&\!\!=\!\!& \ln N \;h_a^{(1)}(\lambda) +
h_a^{(2)}(\lambda,Q^2/\mu_R^2;Q^2/\mu_F^2) + {\cal O}\left(\as(\as
\ln N)^k\right) \,,\\ \label{lnjfun} \ln
J_N^a(\as(\mu_R^2),Q^2/\mu_R^2) &\!\!=\!\!& \ln N \;
f_a^{(1)}(\lambda) + f_a^{(2)}(\lambda,Q^2/\mu_R^2) + {\cal
O}\left(\as(\as \ln N)^k\right) \; , \eeeq where $\lambda=\b0
\as(\mu^2_R) \ln N$. The functions $h^{(1,2)}$ and $f^{(1,2)}$ are
given by
\begin{align}
\label{h1fun}
h_a^{(1)}(\la) =& \f{A_a^{(1)}}{2\pi \b0 \la}
\left[ 2 \la+(1-2 \la)\ln(1-2\la)\right] \;,\\
h_a^{(2)}(\la,Q^2/\mu^2_R;Q^2/\mu_F^2)
=&-\f{A_a^{(2)}}{2\pi^2 \b0^2 } \left[ 2 \la+\ln(1-2\la)\right] -
\f{A_a^{(1)} \gamma_E}{\pi \b0 } \ln(1-2\la)\nn \\
&+ \f{A_a^{(1)} \bone}{2\pi \b0^3}
\left[2 \la+\ln(1-2\la)+\f{1}{2} \ln^2(1-2\la)\right]\nn \\
\label{h2fun}
&+ \f{A_a^{(1)}}{2\pi \b0}\left[2 \la+\ln(1-2\la) \right]
\ln\f{Q^2}{\mu^2_R}-\f{A_a^{(1)}}{\pi \b0} \,\la \ln\f{Q^2}{\mu^2_{F}} \;,
\end{align}
\beeq
\label{fll}
f_a^{(1)}(\lambda) =
&-&\frac{A_a^{(1)}}{2\pi b_0 \lambda}\Bigl[(1-2\lambda)
\ln(1-2\lambda)-2(1-\lambda)
\ln(1-\lambda)\Bigr] \; , \\
\label{fnll}
f_a^{(2)}(\lambda,Q^2/\mu^2_R) =
&-&\frac{A_a^{(1)} b_1}{2\pi b_0^3}\Bigl[\ln(1-2\lambda)
-2\ln(1-\lambda)+\f{1}{2}\ln^2(1-2\lambda)-\ln^2(1-\lambda)\Bigr] \nonumber \\
&+&\frac{B_a^{(1)}}{2\pi b_0}\ln(1-\lambda)
-\frac{A_a^{(1)}\GE}{\pi b_0}\Bigl[\ln(1-\lambda)
-\ln(1-2\lambda)\Bigr] \\
&-&\frac{A_a^{(2)}}{2\pi^2 b_0^2}\Bigl[2\ln(1-\lambda)
-\ln(1-2\lambda)\Bigr]
+ \frac{A_a^{(1)}}{2\pi b_0}\Bigl[2\ln(1-\lambda)
-\ln(1-2\lambda)\Bigr] \ln\frac{Q^2}{\mu^2_R} \; . \nonumber
\eeeq
Here, as before $\b0 = \left( 11 C_A - 2 N_f \right)/12\pi$,
and
\begin{align}
\bone=  \frac{1}{24 \pi^2}
\left( 17 C_A^2 - 5 C_A N_f - 3 C_F N_f \right) \;\; .
\label{bcoef}
\end{align}
We remind the reader that the scale $\mu_F$ represents the
initial-state (final-state) factorization scale $\mu_{FI}$
($\mu_{FF}$) for the radiative factors for the initial (final)
state.  The functions $h^{(1)}$ and $f^{(1)}$ above contain all LL
terms in the perturbative series, while $h^{(2)}$ and $f^{(2)}$
are of NLL accuracy only. For a complete NLL resummation one also
needs the coefficients $\ln\Delta^{{\rm (int)} ab\rightarrow
cd}_{I\, N}$ whose NLL expansion reads: \beeq \label{lndeltams1}
\ln\Delta^{{\rm (int)} ab\rightarrow cd}_{I\,
N}(\as(\mu_R^2),Q^2/\mu_R^2) &\!\!=\!\!& \frac{D_{I\, ab \to c
d}^{(1)}}{2\pi b_0} \;\ln (1-2\lambda) + {\cal O}\left(\as(\as \ln
N)^k\right) \, . \eeeq As we mentioned earlier, the $D_{I\, ab \to
c d}^{(1)}$, and the corresponding ``color weights'' $G_{I\, ab
\to c d}$, are both process and ``color configuration'' dependent.
All the coefficients $D_{I\, ab \to c d}^{(1)}$ and $G_{I\, ab \to
c d}$ that we need to NLL are listed in the Appendix.

\subsection{Coefficients $C_{ab\to cd}^{(1)}$}

We have verified for each subprocess that expansion of the
resummed formulas above to ${\cal O}(\as^3)$ correctly reproduces
the logarithmic terms $\propto \as^3 \ln^2 (N), \,  \as^3 \ln (N)$
known from the full fixed-order
calculations~\cite{Aversa:1988vb,Jager:2002xm}. Comparison
to those calculations also allows to extract the first-order
coefficients $C_{ab\to cd}^{(1)}$. Numerical results for the
coefficients are presented in the Appendix.

\subsection{Matching to the NLO cross section, and inverse Mellin transform}

As we have discussed above, the resummation is achieved in Mellin
moment space. In order to obtain a resummed cross section in
$x_T^2$ space, one needs an inverse Mellin transform. This
requires a prescription for dealing with the singularity
in the perturbative strong coupling constant in
Eqs.~(\ref{Dfct})-(\ref{Dintfct}) or in the
NLL expansions, Eqs.~(\ref{h1fun})-(\ref{lndeltams1}). We will use
the {\em Minimal Prescription} developed in Ref.~\cite{Catani:1996yz},
which relies on use of the NLL expanded forms
Eqs.~(\ref{h1fun})-(\ref{lndeltams1}), and on choosing
a Mellin contour in complex-$N$ space that lies to the {\it left}
of the poles at $\lambda=1/2$ and $\lambda=1$ in the Mellin integrand:
\begin{align}
\label{hadnmin}
\f{p_T^3\, d\sigma^{\rm (res)}(x_T)}{dp_T} &=
\;\int_{C_{MP}-i\infty}^{C_{MP}+i\infty}
\;\frac{dN}{2\pi i} \;\left( x_T^2 \right)^{-N}
\sigma^{\rm (res)}(N) \; ,
\end{align}
where $b_0\as(\mu_R^2)\ln C_{MP}<1/2$, but all other poles
in the integrand are as usual to the left of the contour. The
result defined by the minimal prescription has the property that
its perturbative expansion is an asymptotic series that
has no factorial divergence and therefore
no ``built-in'' power-like ambiguities. Power corrections may
then be added, as phenomenologically required.

When performing the resummation, one of course wants to make full
use of the available fixed-order cross section, which in our case
is NLO (${\cal O}(\as^3)$). Therefore, a matching to this cross
section is appropriate, which may be achieved by expanding the resummed
cross section to ${\cal O}(\as^3)$, subtracting the expanded result
from the resummed one, and adding the full NLO cross section:
\begin{align}
\label{hadnres}
\f{p_T^3\, d\sigma^{\rm (match)}(x_T)}{dp_T} &= \sum_{a,b,c}\,
\;\int_{C_{MP}-i\infty}^{C_{MP}+i\infty}
\;\frac{dN}{2\pi i} \;\left( x_T^2 \right)^{-N+1}
\; f_{a/h_1}(N,\mu_{FI}^2) \; f_{b/h_2}(N,\mu_{FI}^2) \;
D_{c/h}(2N+1,\mu_{FF}^2)
 \nn \\
&\times \left[ \;
\hat{\sigma}^{\rm (res)}_{ab\to cd} (N)
- \left. \hat{\sigma}^{{\rm (res)}}_{ab\to cd} (N)
\right|_{{\cal O}(\as^3)} \, \right]
+\f{p_T^3\, d\sigma^{\rm (NLO)}(x_T)}{dp_T}
 \;\;,
\end{align}
where $\hat{\sigma}^{{\rm (res)}}_{ab\to cd} (N)$ is the resummed cross
section for the partonic channel $ab\to cd$ as given in Eq.~(\ref{eq:res}).
In this way, NLO is taken into account in full, and the soft-gluon
contributions beyond NLO are resummed to NLL. Any double-counting
of perturbative orders is avoided.

\section{Phenomenological Results \label{sec4}}

Starting from Eq.~(\ref{hadnres}), we are now ready to present
some first resummed results at the hadronic level.
This is not meant to be an exhaustive study of
the available data for inclusive-hadron production; rather
we should like to investigate the overall size and relevance
of the resummation effects. We will only consider $\pi^0$ production
in $pp$ collisions and compare to a few selected sets of data.

Let us begin by specifying some ``default'' choices for the
distribution functions that we will use in our studies.
We will use the MRST2002 set of parton densities~\cite{mrst}
and the pion fragmentation functions of~\cite{kkp} (referred to
as ``KKP''). For comparison, we will also present some
results for Kretzer's set~\cite{kretzer} of fragmentation functions.
Note that, according to Eq.(\ref{hadnres}), it is a great
advantage to have parton densities and fragmentation functions
available in moment space. Technically, since the MRST distributions
are not available in moment space, we first performed a fit of a
simple functional form to the MRST distributions, of which we
were then able to take moments. This had to be done separately
for each parton type and for each scale. Concerning the fragmentation
functions, Kretzer's set is anyway set up in moment space, and we
found it possible to analytically take moments of the KKP parameterization.

\begin{figure}[htb]
\begin{center}
\vspace*{-0.6cm}
\epsfig{figure=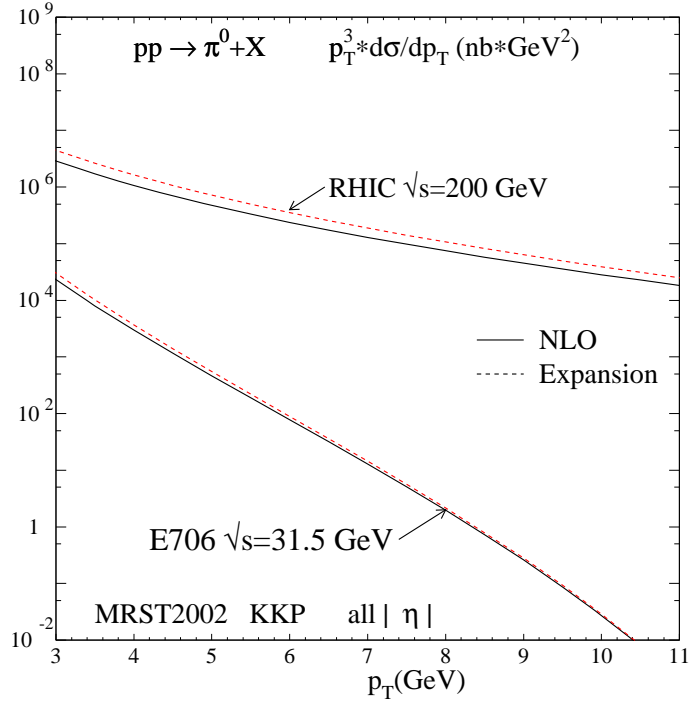,width=0.6\textwidth}
\end{center}
\vspace*{-.5cm}
\caption{Comparison of full NLO cross sections (solid lines) for
$pp\to \pi^0X$ with the NLO (${\cal O}(\as^3)$) expansion of the
resummed cross section (dashed lines), for two different energies.
We have chosen the factorization and renormalization scales
as $p_T$.  \label{fig:softap} }
\vspace*{0.cm}
\end{figure}

As we discussed at the end of subsection~\ref{subsec32}, we generally
expect fairly large effects from soft-gluon resummation for
inclusive-hadron production. This makes it rather important to
be sure that the resummed soft-gluon terms indeed constitute the
dominant part of the cross section and do not, for example,
lead to an overestimate of the higher orders. We therefore
start by identifying the kinematic regions where soft-gluon contributions
are likely to dominate the cross section. A gauge for this is
obtained by comparing the resummed formula expanded to NLO to the full
fixed-order (NLO) perturbative result, that is, by comparing the last two terms
in Eq.~(\ref{hadnres}). Figure~\ref{fig:softap} shows this comparison for
a typical fixed-target energy $\sqrt{S}=31.5$ GeV, and for RHIC's
$\sqrt{S}=200$ GeV. As can be observed, the expansion faithfully
reproduces the NLO result. In the fixed-target regime the agreement
is excellent, except perhaps at the lowest pion transverse momenta,
$p_T\sim 3$ GeV, where the soft approximation tends to yield a slight
overestimate. This is obviously related to the fact that the smaller
$p_T$ (at fixed energy), the further one is away from threshold, so that
the soft-gluon approximations become less reliable. The same is expected to
happen if the energy is increased at fixed transverse momentum. Indeed,
as the curves for $\sqrt{s}=200$~GeV in Fig.~\ref{fig:softap} show,
the NLO-expanded resummed result, while still remarkably close
to the full NLO prediction, gives a less accurate picture of the
latter than at fixed-target energies. Our conclusion from Fig.~\ref{fig:softap}
is therefore that the contributions associated with the near-threshold region
are dominant in the fixed-target regime, implying that resummation will be
relevant and accurate here, even at relatively small transverse momenta.
At colliders, our resummed cross section will likely be too large, and
further improvements in the theoretical framework may be needed\footnote{
It is worth recalling that our matching procedure given by
Eq.~(\ref{hadnres}) ensures that the NLO cross section is always
fully and exactly taken into account in our final ``matched''
cross section, so that any overestimate would only occur at NNLO and beyond.}.
We will briefly return to this point at the end of this paper. In the
following we will primarily focus on the fixed-target regime.

\begin{figure}[htb]
\begin{center}
\vspace*{-0.6cm}
\epsfig{figure=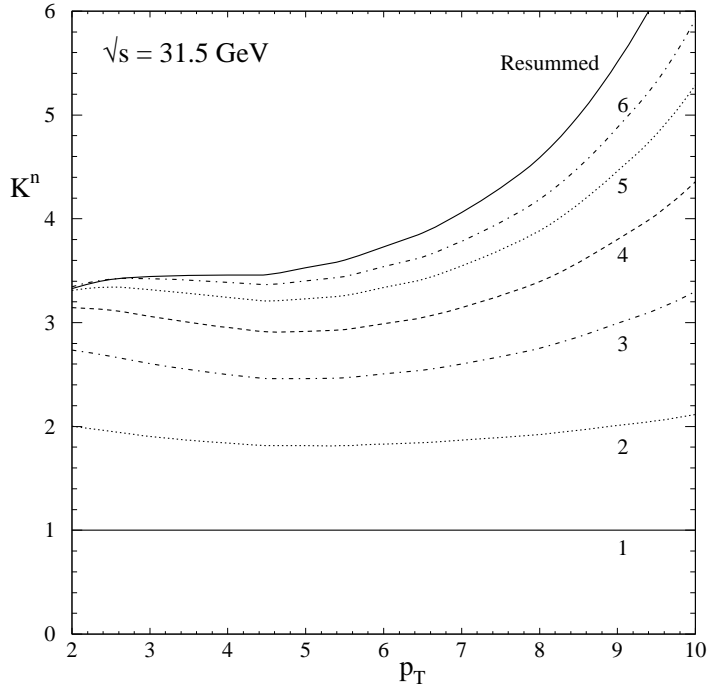,width=0.6\textwidth}
\end{center}
\vspace*{-.5cm}
\caption{``$K$-factors'' relative to NLO as defined in
Eqs.~(\ref{eq:kres}) and~(\ref{ksoftg}) for $pp\to \pi^0X$
in the fixed-target regime. \label{fig:expan}}
\vspace*{1.cm}
\end{figure}

We next investigate how large the higher-order contributions provided
by NLL resummation are. To this end, we go back to Eq.~(\ref{hadnres}) and
take the full resummed result, defined in the minimal prescription and
matched to NLO. As before, the cross sections are integrated over
all rapidities. We define a resummed ``$K$-factor'' as the ratio of the
resummed cross section to the NLO cross section,
\begin{equation}
\label{eq:kres}
K^{{\rm (res)}} = \f{{d\sigma^{\rm (match)}}/{dp_T}}
{{d\sigma^{\rm (NLO)}}/{dp_T}}\, ,
\end{equation}
which is shown for the fixed-target regime, and for scales
$\mu_R=\mu_{FI}=\mu_{FF}=p_T$, by the solid line
in Fig.~\ref{fig:expan}. As can be seen, $K^{{\rm (res)}}$ is very
large, meaning that resummation results in a dramatic enhancement over NLO.
It is then interesting to see how this enhancement builds up
order by order in the resummed cross section. We therefore
expand the matched resummed formula beyond NLO and define the
``soft-gluon $K$-factors''
\begin{equation} \label{ksoftg}
K^n\;\equiv\; \f{{\left. d\sigma^{\rm{(match)}}/{dp_T}\right|_{{\cal O}
(\as^{2+n})}}}{{d\sigma^{\rm (NLO)}}/{dp_T}} \; ,
\end{equation}
which for $n=2,3,\ldots$ give the additional enhancement over full NLO due to
the ${\cal O}(\as^{2+n})$ terms in the resummed formula. Formally,
$K^1=1$ and $K^{\infty}=K^{{\rm (res)}}$ of Eq.~(\ref{eq:kres}). The
results for $K^{2,3,4,5,6}$ are also shown in Fig.~\ref{fig:expan}.
One can see that there are very large large contributions even
beyond NNLO, in particular at the higher $p_T$. Clearly, the
full resummation given by the solid line is required here.

As we have mentioned earlier, we have determined the resummed formulas
for the fully rapidity-integrated cross section, whereas in experiment always
only a certain limited range of rapidity is covered. In order to be able
to compare to data, we therefore approximate the cross section in
the experimentally accessible rapidity region by
\begin{equation}
\f{p_T^3\, d\sigma^{\rm (match)}}{dp_T}({\rm \eta\, in\, exp.\, range})
= K^{{\rm (res)}} \, \f{p_T^3\, d\sigma^{\rm (NLO)}}{dp_T}
({\rm \eta\, in\, exp.\, range})\, ,
\end{equation}
where $K^{{\rm (res)}}$ is as defined in Eq.~(\ref{eq:kres}) in terms
of cross sections integrated over the full region of rapidity. In other
words, we rescale the matched resummed result by the ratio of
NLO cross sections integrated over the experimentally relevant
rapidity region or over all $\eta$, respectively.

\begin{figure}[htb]
\begin{center}
\vspace*{-0.6cm}
\epsfig{figure=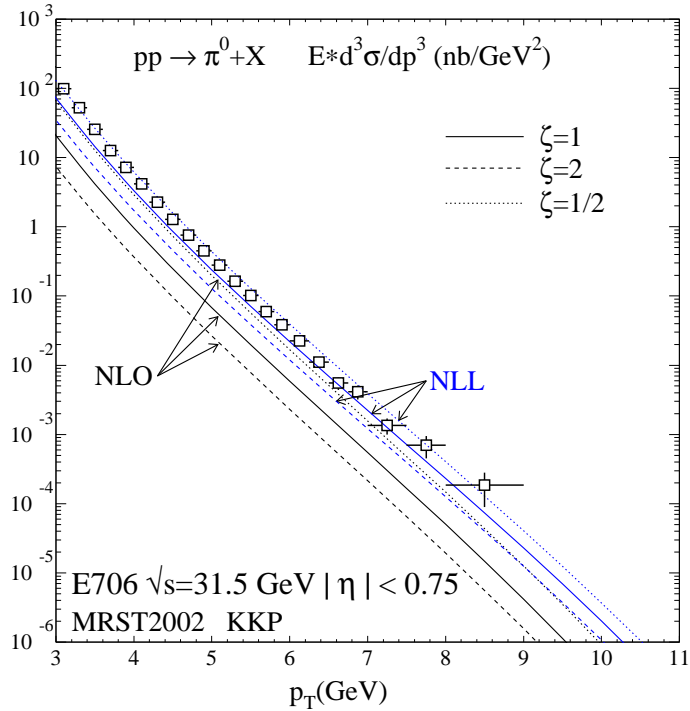,width=0.6\textwidth}
\end{center}
\vspace*{-.5cm}
\caption{NLO and NLL resummed results for the cross
section for $pp\to \pi^0X$ for E706 kinematics. We have
used the KKP fragmentation functions~\cite{kkp}. Results
are given for three different choices of scales,
$\mu_R=\mu_{FI}=\mu_{FF}= \zeta p_T$, where $\zeta=1/2, 1, 2$.
Data are from~\cite{e706}. \label{fig:e706kkp} }
\vspace*{1.cm}
\end{figure}

In Fig.~\ref{fig:e706kkp} we compare the NLL resummed and NLO predictions
to the available data from E706~\cite{e706} for $pp\to \pi^0 X$
at $\sqrt{S}=31.5$~GeV. The data cover $|\eta|<0.75$. We
use the KKP fragmentation functions~\cite{kkp} and give results
for three different choices of scales, $\mu_R=\mu_{FI}=\mu_{FF}=
\zeta p_T$, where $\zeta=1/2, 1, 2$. It is evident that the NLO
result falls far short of the data, which is an observation that
has been made before~\cite{aur,apan,bs}. Furthermore, there is a very
large scale dependence at NLO. The situation is significantly
improved when the NLL resummation is taken into account. As
we already saw in Fig.~\ref{fig:expan}, the NLL matched cross
section is considerably higher than the NLO one, and it shows a
markedly improved comparison to the E706 data, probably satisfactory
in view of the overall uncertainties. Furthermore, the scale dependence
is considerably reduced compared to the NLO calculation, and hence the
accuracy of the prediction is improved.

\begin{figure}[htb]
\begin{center}
\vspace*{-0.6cm}
\epsfig{figure=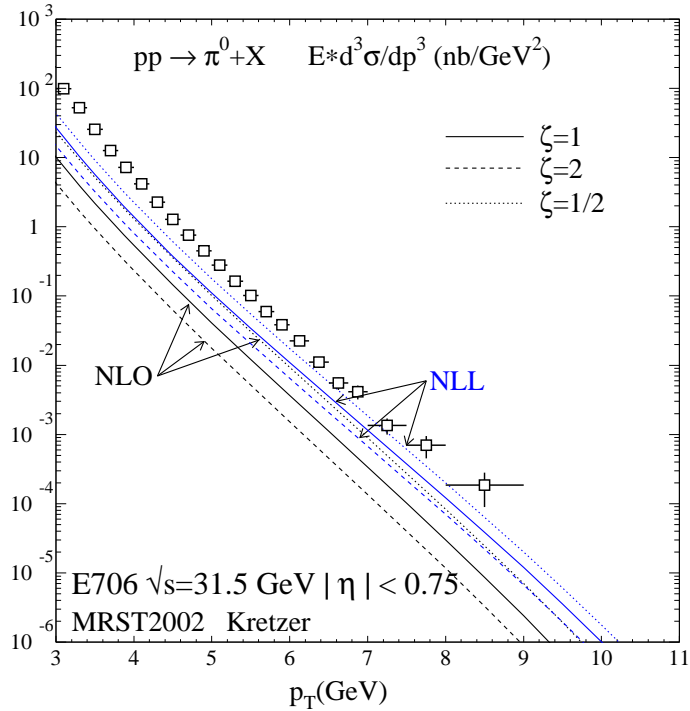,width=0.6\textwidth}
\end{center}
\vspace*{-.5cm}
\caption{Same as Fig.~\ref{fig:e706kkp}, but for the set
of fragmentation functions of~\cite{kretzer}. \label{fig:e706k} }
\vspace*{1.cm}
\end{figure}

Figure~\ref{fig:e706k} shows the same result, but now for the
Kretzer set~\cite{kretzer} of pion fragmentation functions. These
functions are known to be overall significantly smaller than the ones of
KKP, in particular for the gluon fragmentation function which is not
well determined from the $e^+e^-\to hX$ data\footnote{Note that
analyses of hadron production in the additional jet in
$e^+e^-\to b\bar{b}\, {\rm jet}$ events \cite{lepbbg}
do constrain $D_g^{\pi}$ significantly. The $D_g^{\pi}$'s in the
sets of~\cite{kkp} and~\cite{kretzer} are in reasonable agreement with these
data, with the one of~\cite{kretzer} arguably setting a lower
bound on $D_g^{\pi}$.}. One therefore finds that all theory curves
are shifted downward with respect to the results shown in the previous
figure. Nevertheless, the effects due to threshold resummation
remain large.

\begin{figure}[htb]
\begin{center}
\vspace*{-0.6cm}
\epsfig{figure=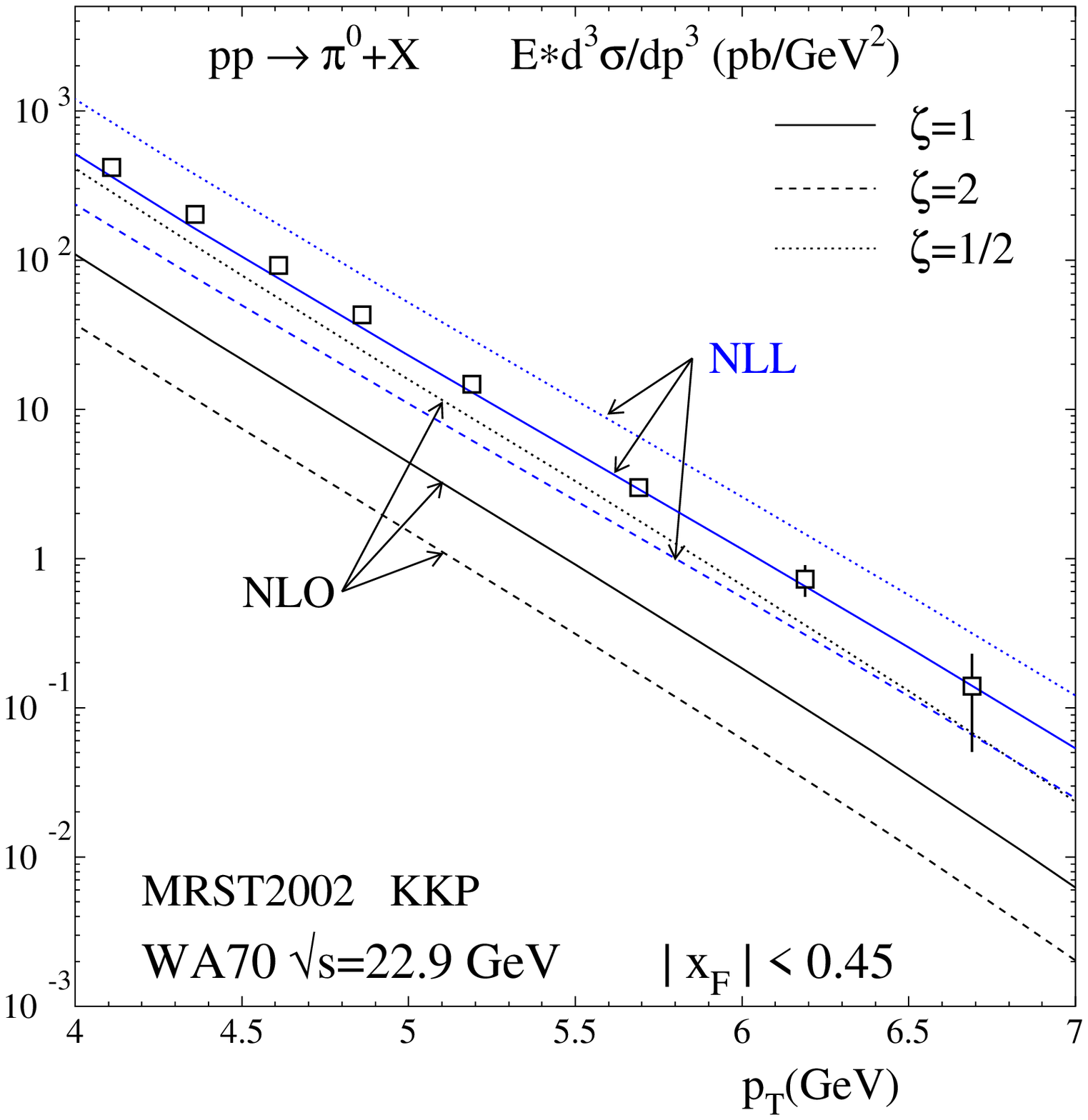,width=0.6\textwidth}
\end{center}
\vspace*{-.5cm}
\caption{Same as Fig.~\ref{fig:e706kkp}, but comparing
to the $pp\to \pi^0 X$ data of WA70~\cite{wa70} at
$\sqrt{S}=22.9$~GeV. \label{fig:wa70kkp}}
\vspace*{1.cm}
\end{figure}

To give another example, Fig.~\ref{fig:wa70kkp} presents a
comparison to data from the WA70 experiment, corresponding to
$\sqrt{S}=31.5$ GeV and $|x_F|<0.45$. Again a significant
enhancement due to NLL resummation is found, resulting in a much
improved agreement between theory and data.

\begin{figure}[htb]
\begin{center}
\vspace*{-0.6cm}
\epsfig{figure=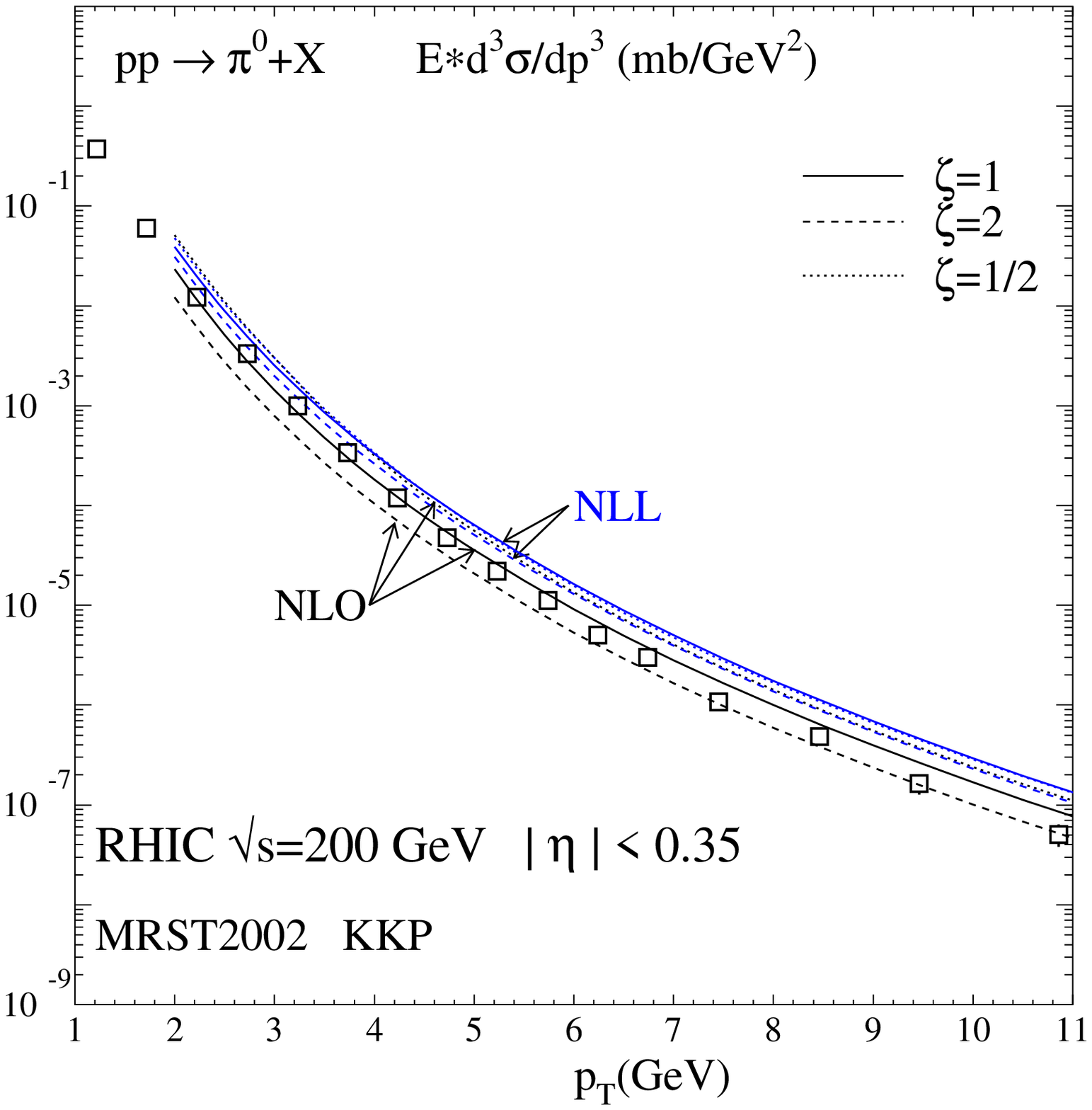,width=0.6\textwidth}
\end{center}
\vspace*{-.5cm}
\caption{Same as Fig.~\ref{fig:e706kkp}, but comparing
to the RHIC PHENIX data of~\cite{phenix}.
\label{fig:rhickkp}}
\vspace*{1.cm}
\end{figure}

Finally, in Fig.~\ref{fig:rhickkp} we repeat the calculations for
the case of proton-proton collisions at RHIC with $\sqrt{S}=200$ GeV and
$|\eta|<0.35$. The data are from the measurement performed by the
PHENIX Collaboration~\cite{phenix}. Again, an enhancement from
resummation is found which however is smaller than in the previous
figures. This is expected since we are further away from
threshold here, due to the much higher energy. Nevertheless,
the enhancement is quite significant at the larger $p_T$,
where in fact the resummed result appears to lie too high.
We emphasize, however, that according to our results shown
in Fig.~\ref{fig:softap} it is likely that the NLL resummation
gives an overestimate of the higher-order corrections in this
case. We therefore do not take the enhancement too literally
and reserve its closer investigation to a future study. We note
that also in this case there is a considerable reduction in the
scale dependence, and that again the fragmentation functions
of~\cite{kretzer} lead to a smaller cross section.

\section{Conclusions and outlook \label{sec5}}

We have studied the NLL all-order resummation of threshold
logarithms in the partonic cross sections relevant for the process
$pp\to h X$ at high transverse momentum of the hadron $h$.
This study has in part been motivated by the observed shortfall
of fixed-order (NLO) cross sections when compared to experimental
data in the fixed-target regime, in contrast with the excellent
agreement of data and theory at colliders. Our numerical results indeed
show a strong enhancement of the cross section over the next-to-leading
order one for typical fixed-target kinematics, significantly improving
the agreement between data and theoretical predictions.
At higher energies, such as at RHIC, the resummation effects are less
important, but more theoretical analysis is needed here due to the
likely relevance of subleading terms.

We emphasize that the contributions generated by
resummation are a well-defined class of higher-order
corrections to the leading-power partonic cross sections that
will be present in the full perturbative series order by
order and actually dominate it. Our results are then also to be
seen in the context of the size of possible non-perturbative
power-suppressed corrections to the cross section.
Any residual shortfall of the resummed theoretical prediction
would need to be attributed to such contributions. In previous
studies, ``intrinsic'' transverse momenta of partons have often
been taken into account in (LO or NLO) calculations of
inclusive-hadron cross sections~\cite{apan,intr}, in order to
bridge the large gaps between data and NLO theory in the
fixed-target regime. These can perhaps best be viewed as
models of the power-suppressed contributions. In the light of
our results, however, much of the enhancements needed for a satisfactory
description of the fixed-target data appears to come from
perturbative contributions, so that power-suppressed contributions
are probably of rather moderate size. It is interesting to
note that resummed perturbation theory itself may provide
information on the structure of power corrections, through
ambiguities in the perturbative series~\cite{pcresu} arising from the
pole in the perturbative running coupling in the expressions
Eqs.~(\ref{Dfct})-(\ref{Dintfct}). A recent study~\cite{gswv}
addressed this issue in the case of single-inclusive cross
sections at large $x_T$ and indeed estimated power corrections
to be not very sizable. On the other hand, it is 
known~\cite{CMN,KO,sv} that
threshold resummation effects are not very large for the prompt
photon cross section in the fixed-target regime, where
discrepancies between data and NLO theory of similar magnitude
as for pion production have been observed in some cases. This issue
clearly needs further study.

We finally emphasize that we regard this study only as the
beginning of a more detailed analysis of threshold resummation for
inclusive-hadron cross sections. There are several points in which
further developments are desirable. First of all, as we noted
earlier, it would be possible in principle (albeit challenging
technically) to perform the resummation correctly for the fully
rapidity-dependent cross section. To do this appears all the more
interesting since it was observed~\cite{bs} that the discrepancies
between NLO and fixed-target data actually increase (at fixed $p_T$) toward
larger rapidities. We expect resummation effects to become even
more important as well at large $\eta$, simply because one is
approaching threshold more closely. It is also possible to improve
the resummation by resumming also terms of the form
$\as^k\ln^{2k-1}(N)/N$ in the partonic cross sections. Such terms
arise from collinear emissions~\cite{higgs,higgsnnll,KSV}; they
are suppressed with respect to the LL and NLL terms but may
nonetheless be of relevance if one is further away from threshold
as for example in the fixed-target regime at lower $p_T$, or a
collider energies. We expect that such additional contributions
would also
further decrease the scale dependence. With these improvements in
place, detailed phenomenological studies of power corrections
might become possible. We finally also note that another important
field for further study would be the effects of threshold
resummation on spin asymmetries, in particular on the
double-longitudinal spin asymmetries $A_{LL}$ measured by
E704~\cite{e704} and at RHIC~\cite{phenix}. We believe that the
significant enhancements due to resummation that we found in this
work strongly motivate all these further studies.

\section*{Acknowledgments}
We are grateful to S.\ Catani and G.\ Sterman for a careful reading
of the manuscript, and to S.\ Kretzer for useful discussions.
The work of D.dF has been partially supported by Conicet,
Fundaci\'on Antorchas, UBACyT and ANPCyT.
W.V.\ is grateful to RIKEN, Brookhaven National Laboratory
and the U.S.\ Department of Energy (contract number DE-AC02-98CH10886) for
providing the facilities essential for the completion of his work.

\section*{Appendix: Results for the various subprocesses}

In this appendix we compile the moment-space expressions
for the Born cross sections for the various partonic
subprocesses, and the process dependent coefficients
$C_{ab\to cd}^{(1)}$, $D_{I\, ab\to cd}$, and $G^{I}_{ab\to cd}$.
Since the $C_{ab\to cd}^{(1)}$ have rather lengthy expressions,
we only give their numerical values for $N_f=5$ and the
 factorization and renormalization scales set to $\mu_{FI}=\mu_{FF}=\mu_R=Q$.

\begin{description}
\item[$qq'\to qq'$:]
\begin{eqnarray}
&&\hat{\sigma}^{{\rm (Born)}}_{qq'\to qq'} (N) =
\frac{\pi C_F}{3C_A} \left(
5 N^2 + 15 N+12\right) B\left(N, \frac{5}{2}\right) \; , \nn \\
&&
G_{1\, qq'\to qq'}=1/3\,,\,\,\,\,\, G_{2\, qq'\to qq'}=2/3\,,
\,\,\,\,\, D^{(1)}_{1\, qq'\to qq'}=-4\, \ln 2\,, \,\,\,\,\,
D^{(1)}_{2\, qq'\to qq'}=0 \; , \nn \\
&& C^{(1)}_{1\, qq'\to qq'}=20.2389 \,\,(N_f=5)\; .
\end{eqnarray}
\item[$q\bar{q'}\to q\bar{q'}$:]
\begin{eqnarray}
&&\hat{\sigma}^{{\rm (Born)}}_{q\bar{q'}\to q\bar{q'}} (N) =
\frac{\pi C_F}{3C_A} \left(
5 N^2 + 15 N+12\right) B\left(N, \frac{5}{2}\right) \; , \nn \\
&&
G_{1\,q\bar{q'}\to q\bar{q'}}=1/9\,,\,\,\,\,\, G_{2\,
q\bar{q'}\to q\bar{q'}}=8/9 \,, \,\,\,\,\, D^{(1)}_{1\,
q\bar{q'}\to q\bar{q'}}=-10/3\, \ln 2\,,\,\,\,\,\, D^{(1)}_{2\,
q\bar{q'}\to q\bar{q'} }=8/3\,
\ln 2  \nn \; , \\
&& C^{(1)}_{1\, q\bar{q'}\to q\bar{q'}}=22.4483 \,\,(N_f=5)\; .
\end{eqnarray}
\item[$q\bar{q}\to q'\bar{q'}$:]
\begin{eqnarray}
&&\hat{\sigma}^{{\rm (Born)}}_{q\bar{q}\to q'\bar{q'}} (N) =
\frac{\pi C_F}{6C_A} \left( N+1\right)\left(N+3\right)
B\left(N+1, \frac{5}{2}\right) \; , \nn \\
&&
G_{1\,q\bar{q}\to q'\bar{q'}}=1\,,\,\,\,\,\,D^{(1)}_{1\,
q\bar{q}\to q'\bar{q'}}=-10/3\, \ln 2\,,\,\,\,\,\,  \nn \; , \\
&& C^{(1)}_{1\, q\bar{q}\to q'\bar{q'}}=7.91881 \,\,(N_f=5)\; .
\end{eqnarray}
\item[$qq\to qq$:]
\begin{eqnarray}
&&\hat{\sigma}^{{\rm (Born)}}_{qq\to qq} (N) =
\frac{2\pi C_F}{3C_A^2} \left(C_A (5 N^2+15N+12) -
2 N (3+2N)\right)B\left(N, \frac{5}{2}\right) \; , \nn \\
&&
G_{1\, qq\to qq}=9/11\,,\,\,\,\,\, G_{2\, qq\to qq}=2/11\,,
\,\,\,\,\, D^{(1)}_{1\, qq\to qq}=-4\, \ln 2\,, \,\,\,\,\,
D^{(1)}_{2\, qq\to qq}=0 \nn \; , \\
&& C^{(1)}_{1\, qq\to qq}=19.535 \,\,(N_f=5)\; .
\end{eqnarray}
\item[$q\bar{q}\to q\bar{q}$:]
\begin{eqnarray}
&&\hat{\sigma}^{{\rm (Born)}}_{q\bar{q}\to q\bar{q}} (N) =
\frac{\pi C_F}{15C_A^2} \left(C_A (11 N^3+59N^2+102 N+60) +
N (N+3)(5+2N)\right)B\left(N, \frac{7}{2}\right) \; , \nn \\
&&
G_{1\,q\bar{q}\to q\bar{q}}=5/21\,,\,\,\,\,\, G_{2\,
q\bar{q}\to q\bar{q}}=16/21 \,, \,\,\,\,\, D^{(1)}_{1\,
q\bar{q}\to q\bar{q}}=-10/3\, \ln 2\,,\,\,\,\,\,
D^{(1)}_{2\,q\bar{q}\to q\bar{q} }=8/3\, \ln 2 \nn \; , \\
&&C^{(1)}_{1\, q\bar{q}\to q\bar{q}}=19.9643 \,\,(N_f=5)\; .
\end{eqnarray}
\item[$q\bar{q}\to gg$:]
\begin{eqnarray}
&&\hat{\sigma}^{{\rm (Born)}}_{q\bar{q}\to gg} (N) =
\frac{\pi C_F}{3C_A} \left(2 C_F(N+2)(5+2N) -
C_A (N+1) (N+3)\right) B\left(N+1, \frac{5}{2}\right) \; , \nn \\
&&
G_{1\,q\bar{q}\to gg}=5/7\,,\,\,\,\,\,
G_{2\, q\bar{q}\to gg}=2/7 \,, \,\,\,\,\,
D^{(1)}_{1\, q\bar{q}\to gg}=-10/3\, \ln 2\,,\,\,\,\,\,
D^{(1)}_{2\,q\bar{q}\to gg}=8/3\, \ln 2  \; , \nn \\
&&C^{(1)}_{1\, q\bar{q}\to gg}=12.4329 \,\,(N_f=5)\; .
\end{eqnarray}
\item[$qg\to qg$:]
\begin{eqnarray}
&&\hat{\sigma}^{{\rm (Born)}}_{qg\to qg} (N) =
\frac{\pi}{6C_A} \left(C_F N (7+5N) +
2C_A (5 N^2+15N+12)\right) B\left(N, \frac{5}{2}\right) \; , \nn \\
&&
G_{1\,qg\to qg}=45/88\,,\,\,\,\,\, G_{2\, qg\to qg}=25/88 \,, \,\,\,\,\,
G_{3\, qg\to qg}=18/88\,,\,\,\,\,\,  \nn \\
&&D^{(1)}_{1\, qg\to qg}=-14/3\, \ln 2\,,\,\,\,\,\,
D^{(1)}_{2\,qg\to qg}=10/3\, \ln 2  \,,\,\,\,\,\,
D^{(1)}_{3\,qg\to qg}=-2/3\, \ln 2 \; , \nn \\
&&C^{(1)}_{1\, qg\to qg}=15.4167 \,\,(N_f=5)\; .
\end{eqnarray}
\item[$qg\to gq$:]
\begin{eqnarray}
&&\hat{\sigma}^{{\rm (Born)}}_{qg\to gq} (N) =
\frac{\pi}{6C_A} \left(C_F N (7+5N) +
2C_A (5 N^2+15N+12)\right) B\left(N, \frac{5}{2}\right) \; , \nn \\
&&
G_{1\,qg\to gq}=45/88\,,\,\,\,\,\, G_{2\, qg\to gq}=25/88 \,, \,\,\,\,\,
G_{3\, qg\to gq}=18/88 \,,\,\,\,\,\,  \nn \\
&&D^{(1)}_{1\, qg\to gq}=-8\, \ln 2\,,\,\,\,\,\,
D^{(1)}_{2\,qg\to gq}=0  \,,\,\,\,\,\,
D^{(1)}_{3\,qg\to gq}=-4\, \ln 2 \; , \nn \\
&&C^{(1)}_{1\, qg\to gq}=22.4474 \,\,(N_f=5)\; .
\end{eqnarray}
\item[$gg\to gg$:]
\begin{eqnarray}
&&\hat{\sigma}^{{\rm (Born)}}_{gg\to gg} (N) =
\frac{\pi C_A}{5C_F} \left(9 N^3 +45 N^2 + 72 N + 40
\right) B\left(N, \frac{7}{2}\right) \; , \nn \\
&&
G_{1\,gg\to gg}=1/3\,,\,\,\,\,\, G_{2\, gg\to gg}=1/2 \,, \,\,\,\,\,
G_{3\, gg\to gg}=1/6\,,\,\,\,\,\, \nn \\
&&D^{(1)}_{1\, gg\to gg}=0\,,\,\,\,\,\,
D^{(1)}_{2\,gg\to gg}=-10\, \ln 2  \,,\,\,\,\,\,
D^{(1)}_{3\,gg\to gg}=6\, \ln 2  \; , \nn \\
&&C^{(1)}_{1\, gg\to gg}=21.1977 \,\,(N_f=5)\; .
\end{eqnarray}
\item[$gg\to q\bar{q}$:]
\begin{eqnarray}
&&\hat{\sigma}^{{\rm (Born)}}_{gg\to q\bar{q}} (N) =
\frac{\pi}{12C_A C_F} \left(2 C_F(N+2)(5+2N) -
C_A (N+1) (N+3)\right) B\left(N+1, \frac{5}{2}\right) \; , \nn \\
&&
G_{1\,gg\to q\bar{q}}=5/7\,,\,\,\,\,\,
G_{2\, gg\to q\bar{q}}=2/7 \,, \,\,\,\,\,
D^{(1)}_{1\, gg\to q\bar{q}}=0\,,\,\,\,\,\,
D^{(1)}_{2\,gg\to q\bar{q} }=6\, \ln 2  \; , \nn \\
&&C^{(1)}_{1\, gg\to q\bar{q}}=16.7962 \,\,(N_f=5)\; .
\end{eqnarray}
\end{description}
In the above expressions, $B(a,b)$ is the Beta-function.


\begin{thebibliography}{99}

\bibitem{old} B.~L.~Combridge, J.~Kripfganz and J.~Ranft,
Phys.\ Lett.\ B {\bf 70}, 234 (1977); \\
J.~F.~Owens, E.~Reya and M.~Gl\"{u}ck, Phys.\ Rev.\ D {\bf 18}, 1501 (1978).

\bibitem{Aversa:1988vb}
F.~Aversa, P.~Chiappetta, M.~Greco and J.~P.~Guillet,
Nucl.\ Phys.\ B {\bf 327} (1989) 105.

\bibitem{ddf} D.\ de Florian, Phys. Rev. {\bf D67}, 054004 (2003)
[arXiv:hep-ph/0210442].

\bibitem{Jager:2002xm}
B.\ J\"{a}ger, A.\ Sch\"{a}fer, M.\ Stratmann,
and W.\ Vogelsang, Phys. Rev. {\bf D67}, 054005 (2003)
[arXiv:hep-ph/0211007].

\bibitem{aur} P.~Aurenche, M.~Fontannaz, J.~P.~Guillet, B.~A.~Kniehl,
and M.~Werlen, Eur.\ Phys.\ J.\ C {\bf 13}, 347 (2000) [arXiv:hep-ph/9910252].

\bibitem{apan} U.~Baur {\it et al.}, arXiv:hep-ph/0005226.

\bibitem{bs} C.~Bourrely and J.~Soffer,
Eur.\ Phys.\ J.\ C {\bf 36}, 371 (2004) [arXiv:hep-ph/0311110].

\bibitem{kkp1} B.~A.~Kniehl, G.~Kramer and B.~Potter,
Nucl.\ Phys.\ B {\bf 597}, 337 (2001) [arXiv:hep-ph/0011155].

\bibitem{phenix} S.~S.~Adler {\it et al.}  [PHENIX Collaboration],
Phys.\ Rev.\ Lett.\  {\bf 91}, 241803 (2003) [arXiv:hep-ex/0304038].

\bibitem{star} J.~Adams {\it et al.}  [STAR Collaboration],
Phys.\ Rev.\ Lett.\  {\bf 92}, 171801 (2004) [arXiv:hep-ex/0310058].

\bibitem{dyresum} G.~Sterman, Nucl.\ Phys.\ B {\bf 281}, 310 (1987);\\
S.~Catani and L.~Trentadue, Nucl.\ Phys.\ B {\bf 327}, 323 (1989);
Nucl.\ Phys.\ B {\bf 353}, 183 (1991).

\bibitem{KS}
N.~Kidonakis and G.~Sterman,
Nucl.\ Phys.\ B {\bf 505}, 321 (1997)
[arXiv:hep-ph/9705234].

\bibitem{BCMN} R.~Bonciani, S.~Catani, M.~L.~Mangano and P.~Nason,
Phys.\ Lett.\ B {\bf 575}, 268 (2003) [arXiv:hep-ph/0307035].

\bibitem{LOS} E.~Laenen, G.~Oderda and G.~Sterman, Phys.\ Lett.\ B
{\bf 438}, 173 (1998) [arXiv:hep-ph/9806467].

\bibitem{CMN} S.~Catani, M.~L.~Mangano and P.~Nason, JHEP {\bf 9807},
024 (1998) [arXiv:hep-ph/9806484]; \\
S.~Catani, M.~L.~Mangano, P.~Nason, C.~Oleari and W.~Vogelsang,
JHEP {\bf 9903}, 025 (1999) [arXiv:hep-ph/9903436].

\bibitem{KO} N.~Kidonakis and J.~F.~Owens,
Phys.\ Rev.\ D {\bf 61}, 094004 (2000) [arXiv:hep-ph/9912388].

\bibitem{sv} G.~Sterman and W.~Vogelsang,
JHEP {\bf 0102}, 016 (2001) [arXiv:hep-ph/0011289].

\bibitem{KOS} N.~Kidonakis, G.~Oderda and G.~Sterman,
Nucl.\ Phys.\ B {\bf 525}, 299 (1998) [arXiv:hep-ph/9801268];
Nucl.\ Phys.\ B {\bf 531}, 365 (1998) [arXiv:hep-ph/9803241].

\bibitem{KO1} N.~Kidonakis and J.~F.~Owens, Phys.\ Rev.\ D {\bf 63},
054019 (2001) [arXiv:hep-ph/0007268].

\bibitem{higgs} M.~Kr\"{a}mer, E.~Laenen and M.~Spira,
Nucl.\ Phys.\ B {\bf 511}, 523 (1998)
[arXiv:hep-ph/9611272].

\bibitem{higgsnnll}
S.~Catani, D.~de Florian, M.~Grazzini and P.~Nason,
JHEP {\bf 0307}, 028 (2003) [arXiv:hep-ph/0306211].

\bibitem{cc} M.~Cacciari and S.~Catani, Nucl.\ Phys.\ B {\bf 617},
253 (2001) [arXiv:hep-ph/0107138].

\bibitem{KT} J.~Kodaira and L.~Trentadue,
Phys.\ Lett.\ B {\bf 112}, 66 (1982); Phys.\ Lett.\ B {\bf 123},
335 (1983);\\ S.~Catani, E.~D'Emilio and L.~Trentadue,
Phys.\ Lett.\ B {\bf 211}, 335 (1988).

\bibitem{Catani:1996yz} S.~Catani, M.~L.~Mangano, P.~Nason
and L.~Trentadue, Nucl.\ Phys.\ B {\bf 478}, 273 (1996)
[arXiv:hep-ph/9604351].

\bibitem{mrst} A.~D.~Martin, R.~G.~Roberts, W.~J.~Stirling and R.~S.~Thorne,
Eur.\ Phys.\ J.\ C {\bf 28}, 455 (2003) [arXiv:hep-ph/0211080].

\bibitem{kkp} B.~A.~Kniehl, G.~Kramer and B.~Potter,
Nucl.\ Phys.\ B {\bf 582}, 514 (2000) [arXiv:hep-ph/0010289].

\bibitem{kretzer} S.~Kretzer, Phys.\ Rev.\ D {\bf 62}, 054001 (2000)
[arXiv:hep-ph/0003177].

\bibitem{e706}
L.~Apanasevich {\it et al.}  [Fermilab E706 Collaboration],
Phys.\ Rev.\ D {\bf 68}, 052001 (2003) [arXiv:hep-ex/0204031].

\bibitem{lepbbg} See, e.g.: P.~Abreu {\it et al.}  [DELPHI Collaboration],
Eur.\ Phys.\ J.\ C {\bf 13}, 573 (2000), and references therein.

\bibitem{wa70}
M.~Bonesini {\it et al.}  [WA70 Collaboration],
Z.\ Phys.\ C {\bf 38}, 371 (1988).

\bibitem{intr} A.~P.~Contogouris, R.~Gaskell and S.~Papadopoulos,
Phys.\ Rev.\ D {\bf 17}, 2314 (1978); \\
L.~Apanasevich {\it et al.}, Phys.\ Rev.\ D {\bf 59},
074007 (1999) [arXiv:hep-ph/9808467]; \\
U.~D'Alesio and F.~Murgia, Phys.\ Rev.\ D {\bf 70}, 074009 (2004)
[arXiv:hep-ph/0408092].

\bibitem{pcresu} G.~'t Hooft, Nucl.\ Phys.\ B {\bf 138}, 1 (1978); \\
A.~H.~Mueller, Nucl.\ Phys.\ B {\bf 250}, 327 (1985); \\
M.~Beneke and V.~M.~Braun, arXiv:hep-ph/0010208, and references
therein; \\
M.~Beneke and V.~M.~Braun,
Nucl.\ Phys.\ B {\bf 454}, 253 (1995) [arXiv:hep-ph/9506452]; \\
Y.~L.~Dokshitzer, G.~Marchesini and B.~R.~Webber,
Nucl.\ Phys.\ B {\bf 469}, 93 (1996) [arXiv:hep-ph/9512336];
JHEP {\bf 9907}, 012 (1999) [arXiv:hep-ph/9905339]; \\
G.~P.~Korchemsky and G.~Sterman, Nucl.\ Phys.\ B {\bf
437}, 415 (1995) [arXiv:hep-ph/9411211]; \\
R.~Akhoury and V.~I.~Zakharov, Phys.\ Lett.\ B {\bf 357}, 646 (1995)
[arXiv:hep-ph/9504248]; Nucl.\ Phys.\ B {\bf 465}, 295 (1996)
[arXiv:hep-ph/9507253]; Phys.\ Rev.\ Lett.\  {\bf 76}, 2238 (1996)
[arXiv:hep-ph/9512433]; \\
G.~Sterman and W.~Vogelsang, arXiv:hep-ph/9910371.

\bibitem{gswv} G.~Sterman and W.~Vogelsang,
Phys.\ Rev.\ D {\bf 71}, 014013 (2005) [arXiv:hep-ph/0409234].

\bibitem{KSV} A.~Kulesza, G.~Sterman and W.~Vogelsang,
Phys.\ Rev.\ D {\bf 66}, 014011 (2002) [arXiv:hep-ph/0202251].

\bibitem{e704} D.~L.~Adams {\it et al.}  [E581 Collaboration],
Phys.\ Lett.\ B {\bf 261}, 197 (1991); \\
D.~L.~Adams {\it et al.}  [FNAL E581/704 Collaboration],
Phys.\ Lett.\ B {\bf 336}, 269 (1994).

\end{thebibliography}
\end{document}